\documentclass[%
 reprint,
superscriptaddress,
 amsmath,amssymb,
 aps,pre,hyphens,floatfix
]{revtex4-2}

\usepackage[]{amsmath}
\usepackage{mathrsfs}
\usepackage{mathtools} 
\usepackage{enumerate}
\usepackage{caption}
\usepackage[usenames,dvipsnames,svgnames,table]{xcolor}
\usepackage{IEEEtrantools}
\usepackage{graphicx}% Include figure files
\usepackage{subcaption}
\usepackage{ragged2e} % for justifying
\usepackage{array}
\usepackage{multirow}
\usepackage{verbatim}
\usepackage{soul}
\usepackage{tabularx}
\usepackage{booktabs,siunitx}
\usepackage[T1]{fontenc}
\usepackage{hyperref}
\usepackage{physics}
\usepackage{siunitx}
\usepackage{makecell}
\usepackage{float}

\newcommand{\mycomment}[1]{}

\newcommand{\av}[1]{\langle {#1} \rangle}
\newcommand{\stk}[1]{\ifmmode\text{\sout{\ensuremath{#1}}}\else\sout{#1}\fi}

\AtBeginDocument{\RenewCommandCopy\qty\SI}

\usepackage{setspace}

\usepackage{nomencl}
\makenomenclature

\begin{document}

\title{Quantitative comparison of power grid reinforcements}
% Force line breaks with \\ 

\author{B\'{a}lint Hartmann}
\email[]{hartmann.balint@ek.hun-ren.hu}
\affiliation{Institute of Energy Security and Environmental Safety, HUN-REN Centre for Energy Research, P.O. Box 49, H-1525 Budapest, Hungary}

\author{G\'{e}za \'{O}dor}
%\email[]{odor@mfa.kfki.hu}
\affiliation{Institute of Technical Physics and Materials Science, HUN-REN Centre for Energy Research, P.O. Box 49, H-1525 Budapest, Hungary}

\author{Krist\'of Benedek}
%\email[]{buzasbesenyo@gmail.com}
\affiliation{Institute of Technical Physics and Materials Science, HUN-REN Centre for Energy Research, P.O. Box 49, H-1525 Budapest, Hungary}
\affiliation{Budapest University of Technology and Economics, Műegyetem rkp. 3, H-1111 Budapest, Hungary}

\author{Istv\'an Papp}
%\email[]{istvan.papp@ek-cer.hu}
\affiliation{Institute of Technical Physics and Materials Science, HUN-REN Centre for Energy Research, P.O. Box 49, H-1525 Budapest, Hungary}
\affiliation{HUN-REN  Wigner Research Centre for Physics, P.O. Box 49, H-1525 Budapest, Hungary, Hungary}

\author{Michelle T. Cirunay}
\affiliation{Institute of Technical Physics and Materials Science, HUN-REN Centre for Energy Research, P.O. Box 49, H-1525 Budapest, Hungary}

\date{\today}
% It is always \today, today, but any date may be explicitly specified

\begin{abstract}
This paper presents a quantitative comparison of power grid reinforcement strategies.  We evaluate three approaches: (1) doubling transmission links (bridges) between different communities, (2) adding bypasses around weakly synchronized nodes, and (3) reinforcing edges that trigger the largest cascade failures.
We use two different models of the Hungarian high-voltage network. These models are built from the official data provision of the transmission system operator, thus eliminating the assumption typically used by other studies. The coupling strength distribution of the Hungarian models shows good agreement with our previous works using the European and the North-American grids.

Additionally, we examine the occurrence of Braess’ paradox, where added transmission capacity unexpectedly reduces overall stability. Our results show that reinforcement through community-based bridge duplication yields the most significant improvements across all parameters. A visual comparison highlights differences between this method and traditional reinforcement approaches. To the authors knowledge, this is the first attempt to quantitatively compare results of oscillator-based studies and that relying on power system analysis software.

Characteristic results of line-cut simulations reveal cascade size distributions with fat-tailed decays for medium coupling strengths, while exponential behavior emerges for small and large couplings. The observed exponents are reminiscent to the continuously changing exponents by Griffiths effects near to a hybrid type of phase transition.

\end{abstract}

\maketitle

\section{Introduction\label{sec:1}}
Due to their increasing complexity and interconnectedness, modern power systems became more vulnerable to various disturbances, including equipment failures, cyber-physical attacks, and extreme weather events, just to name a few \cite{Panteli2015,Paul2022,Avraam2023,Rajkumar2023,Diaba2024,Hawker2024,Karagiannakis2024}. Ensuring the resilience of power systems requires a systematic approach to reinforcing their topology and mitigating the risk of cascading failures. However, various grid reinforcement methodologies tend to serve different purpose; some more theoretical, some more practical.

Traditional reinforcement methods, such as those used in long-term infrastructure planning (e.g. the Ten-Year Network Development Plan by ENTSO-E) rely on contingency analysis and cost-benefit evaluations as a guidance for installing new infrastructure. These traditional approaches may struggle to account for dynamic stability and rarely consider emerging threats. In contrast, alternative methodologies rooted in synchronization theory, such as the Kuramoto model and oscillator-based techniques, offer new insights into grid stability by analyzing phase coherence. However, the models used by this approach are usually simplified and homogeneous. A third group of methods is rooted in graph-theoretical principles, using community detection and bridging line placement to optimize power grid modularity and enhance resilience.

Alongside these approaches, optimization  (mathematical, heuristic, metaheuristic), machine learning models, and cyber-physical resilience frameworks contribute to this field. The present paper does not intend to provide a comprehensive overview of those studies, but recommends some important papers \cite{1198335,Hemmati2013,Hemmati2013b,8482504,10804778}.

The primary aim of the paper is to present a comparative analysis of traditional, Kuramoto-based and community-based methodologies, highlighting their advantages, limitations. To our knowledge, this is the first attempt to quanitatively compare those methods, using a highly precise grid model.

The structure of the paper is as follows. Section II. presents the Hungarian high-voltage network model that was used for the studies, and the various reinforcement methods. Section III. presents and compares the results. Section IV. discusses the results in light of the Braess's paradox, while conclusions are drawn in Section V.

\begin{table}
    \centering 
    \resizebox{\columnwidth}{!}{
    \begin{tabular}{|p{0.22\columnwidth}|p{0.25\columnwidth}|p{0.2\columnwidth}|p{0.2\columnwidth}|p{0.15\columnwidth}|}
        \hline
        \textbf{Methods} & \textbf{Techniques} & \textbf{Pros} & \textbf{Cons} & \textbf{Ref.} \\
        \hline
        Traditional methods & N-1 security, contingency, cost-benefit analysis, reliability indices & Well-established, regulatory-backed, cost-effective & Limited in capturing dynamic stability, and emerging risks & \cite{entsoeEntsoePlanning,1198335,4349068,Hemmati2013,Hemmati2013b,7954668,8482504,Byles2023,10804778} \\
        \hline
        Kuramoto-based approaches & Phase synchronization analysis, coherence-based methods & Captures dynamic stability, identifies weak synchronization areas & Simplified grid model, may not fully capture transient behavior & \cite{Krakowski2017,Mazauric2018,TurkishPG,Lacerda2021,Zou2024,PhysRevResearch.6.013194} \\
        \hline
        Community-based approaches & Graph clustering, bridging line placement, entropy-based resilience metrics & Improves resilience by preventing cascading failures, enhances inter-area connectivity, modular approach & Requires detailed topological data, effectiveness depends on accurate detection algorithms & \cite{Wang2019,Guerrero2019,8663432,9895255,PhysRevResearch.6.013194} \\
        \hline
    \end{tabular}
    }
    \caption{Comparison of power grid reinforcement methodologies}
    \label{table:comparison}
\end{table}

\section{Models and methods\label{sec:2}}
In \cite{HU387cikk} we investigated 12 different levels of electrical parameter approximations of the
Hungarian high-voltage power grid. These models were built using the official data provision of the Hungarian transmission system operator, including electrical and topological parameters of the grid and measurements on nodal behavior. The use of a model with such high precision is a cornerstone for performing the quantitative analyses presented in this paper.
In \cite{HU387cikk}, we checked in detail how synchronization is influenced on the same structural object but with different levels of capturing reality via the parametrization. Now, we narrow our selection two of the most heterogeneous scenarios; \#8, in which coupling strength is calculated using the actual thermal capacity limit of the power line, and \#12, where coupling strength is determined by the admittance.
Doing so, we will implement different grid reinforcement algorithms and then study their effects both regarding synchronization and cascade failures. With this, we present the counterpart of our aforementioned analysis, as now the model for parametrization will be fixed and the structural object will be changed with each reinforcement method.

\subsection{Network model\label{sec:2A}}
\begin{figure}[H]
     \centering    \includegraphics[width=\columnwidth ]{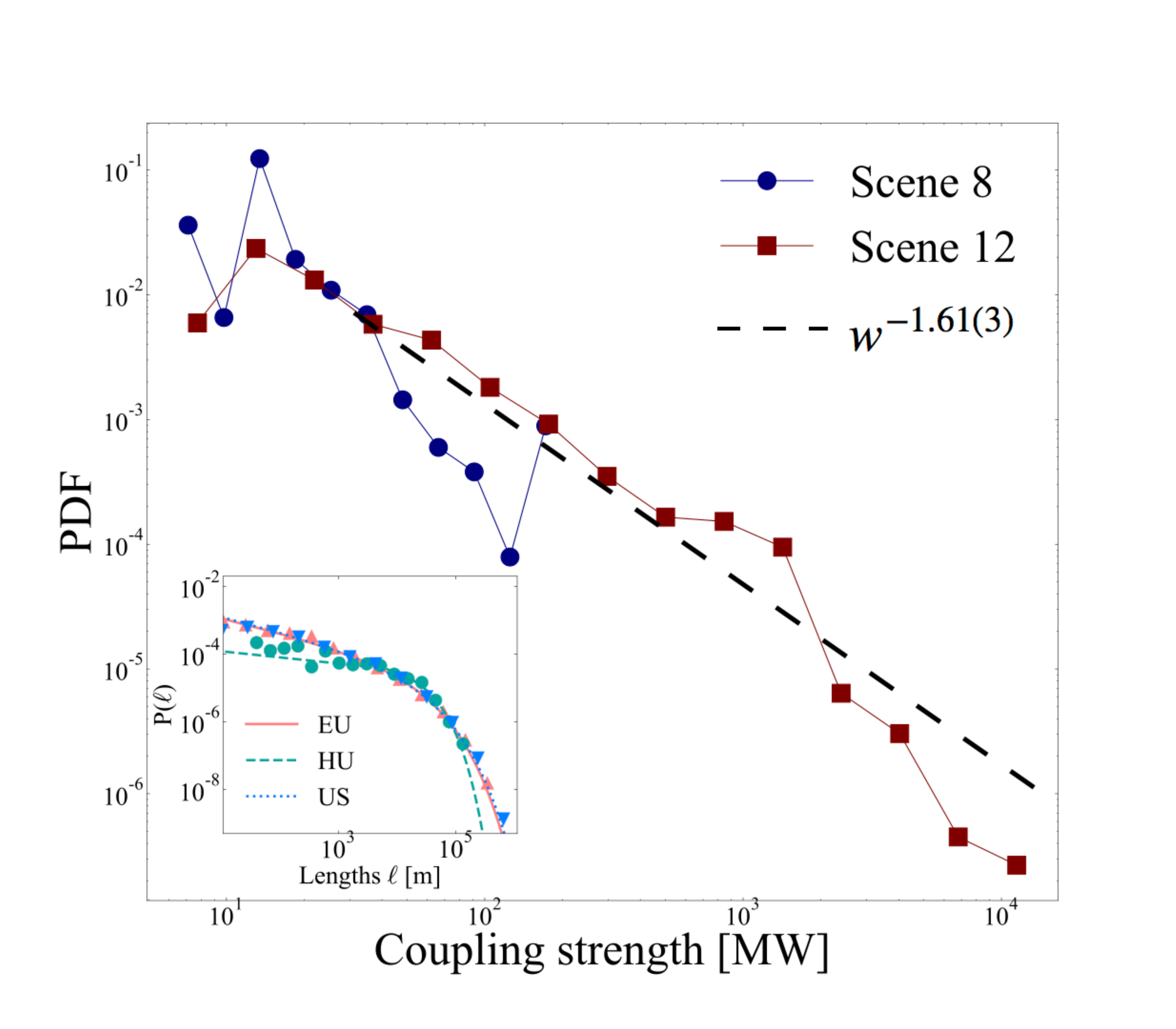}
     \caption{Global link weight distributions of scenarios \#8 (blue circles) and \#12 (red squares). The dashed line shows a PL fit for the tail of scenario \#12, for $w \geq 40$ MW. Inset: Length distributions for the EU ($x_0 = 9867.72, c = 0.555$), Hungary ($x_0 = 17450.72, c = 0.861$), and US ($x_0 = 9720.26, c = 0.536$) power grids fitted with stretched exponentials.}\label{fig:cascades_weight_length}
 \end{figure}

Figure~\ref{fig:cascades_weight_length} shows the link weight distributions of the Hungarian high-voltage grid for both scenarios. The graph edge weights, expressed by coupling strengths, are calculated from the thermal capacity limit and specific impedance values as described in~\cite{HU387cikk}. In scenario \#8 the link length is not accounted for in the weight calculation while the opposite is true in the case of scenario \#12. Looking at the weight distributions, we observe how the effect of doing so as the link weights distribution for scenario \#12  scales according to a PL with an exponent of approximately $-1.6(3)$. On the other hand, this is not observed in the case of scenario \#8.

As accounting for the lengths in scenario \#12 seemed to cause the PL behavior in its coupling strength (weight) distribution, in the inset of Figure~\ref{fig:cascades_weight_length} we show the length distributions of the EU, HU, and the US powergrids. Interestingly, neither of them follow a PL fit but instead a stretched exponential of the form $P(x) \sim e^{-(\frac{x}{x_0})^c}$ with the fitting parameters $x_0$ and $c$ for each dataset shown in the caption. Stretched exponentials are fat-tailed distributions with characteristic scales often proposed as alternative to PL~\cite{LaherrereEPJB1998}. As power losses and voltage drop are proportional to power line lengths, this may explain why we cannot observe a PL behavior (where extreme lengths are possible) in the lengths distributions of the datasets being considered. Rather, a characteristic length is observed which can represent optimal cable length for efficient power transmission.

\subsection{Synchronization model of the power grid\label{sec:2B}}
As discussed in our previous work~\cite{HU387cikk}, one model to simulate power system dynamics is via the so-called swing equations~\cite{swing}, which corresponds mathematically to a set of second-order Kuramoto-equations~\cite{fila}. To make our results comparable with previous works~\cite{Taher_2019,POWcikk,USAEUPowcikk,HARTMANN2024101491} describing the dynamics of a network of $N$ oscillators with phases $\theta_i(t)$ and $\omega_i=\dot{\theta_i}$ we used the specific form
\begin{equation}\label{eq:kur2_phys}
    \dot{\omega}_i = -\frac{D_i \omega_i}{M_i \omega_S} + \frac{L_i}{M_i \omega_S} 
    + \lambda \sum_{j=1}^{N} \frac{Y_{ij} V_i V_j}{M_i \omega_S} \sin(\theta_j - \theta_i) \ .
\end{equation}
Here $D_i$ describes the damping effect of element $i$ in the system with physical dimension $\left[\frac{\si{\kg\cdot\m^2}}{\si{\s}^2}\right]$, 
$L_i$ $\left[\frac{\si{\kg\cdot\m^2}}{\si{\s}^3}\right]$ is the power capacity, 
$Y_{ij}=\frac{1}{X_{ij}}$ $\left[\frac{1}{\si{\ohm}}\right]$ is the susceptance (inverse of reactance) of lines, 
$V_i$ $[\si{\volt}]$ is the  voltage level, $M_i$ $\left[\si{\kg\cdot\m^2}\right]$ is the moment of inertia and 
$\lambda$ is the fraction of total transmitted power in the gird with respect to the maximum. Furthermore, $\omega_S$ is the system frequency and we also have an intrinsic frequency of nodes $\Omega_i = 50$ Hz (in Europe), which can be transformed out in a rotating frame, thus we have omitted it in the calculations. Our frequency results show the deviations from this value. To solve these equations numerically, the time step resolution of the calculations was set to be $\Delta t= 0.25\;\si{\s}$.

In order to model nodal power fluctuations, we have added a multiplicative, quenched term to Eq.~\ref{eq:kur2_phys} as
\begin{equation}\label{eq:noise}
\eta_{in} = 0.05 \xi_n \frac{L_i}{M_i \omega_S}
\end{equation}
where $\xi_n \in N(0,1)$ is a random variable, drawn from a zero-centered Gaussian distribution. The value $0.05$ was chosen by an assumption of $5\%$ fluctuation amplitude of the energy sources and sinks. To solve the set of differential equations we used the adaptive Runge-Kutta-Fehlberg 
method~\cite{watts_shampine_burkardt_r8_rkf45_2004} from the package Numerical Recipies~\cite{Press2007}.

To quantify synchronization we measured the Kuramoto order parameter of phases 
\begin{equation}\label{ordp}
z(t_k) = r(t_k) \exp\left[i \theta(t_k)\right] = 1 / N \sum_j \exp\left[i \theta_j(t_k)\right]
\end{equation}
and the frequency spread: 
\begin{equation}\label{FOP}
        \Omega(t_k) = \frac{1}{N} \langle \sum_{j=1}^N (\overline\omega(t_k)-\omega_jt_k))^2 \rangle
\end{equation}
where $\overline\omega(t_k)$ denotes the mean frequency within each respective sample at the $k$-th time step: $t_k = 1 + 1.08^{k}$. Averages and histograms were calculated from $640 \le  n \le 3000$ independent samples of different initial fluctuations
\begin{equation}\label{KOP}
R(t_k) = \langle r(t_k)\rangle
\end{equation}
as well as for the variance of the frequencies.

Furthermore, the following quantity, describing the energy of classical rotator models, 
suggested by~\cite{schroder2017} and has recently been tested by us for the Hungarian power grid~\cite{HU387cikk}:
\begin{equation}
    r_{uni}(t_k) = 1/(\sum_{i,j}^N w_{ij}) \sum_{i,j}^N w_{ij} \cos(\theta_i-\theta_j) 
\end{equation}
and its sample and temporal average in the steady state:
\begin{equation}\label{eq:runi}
    R_{uni}=\langle r_{uni}(t_k) \rangle
\end{equation}
Fluctuations of the order parameter are also measured via the standard deviations of sample and temporal averages in the steady state.

Following relaxation to the steady state started from phase synchronized or randomized $\theta_i(0)$ values, we introduced single line cuts to initiate power-failure cascades. Further line cuts are committed if the condition 
\begin{equation}
\vert \sin(\theta_i(t)-\theta_j(t)) \vert  > T    
\end{equation}
is satisfied between connected nodes, where $T$ describes the transmission capacity of the power lines~\cite{Crucitti_2004,TurkishPG,Schäfer2018}.
In such cases, these edges are removed from the adjacency matrix, by setting $Y_{ij}=0$. We followed the number of failed edges in this so-called 'line-cut' phase of our simulations up to several thousands of iteration steps, until the end of the cascades. 

\subsection{Strengthening the network via detecting communities - bridges\label{sec:2C}}
Detecting communities in networks aims to identify groups of nodes in the network that are more densely connected to each other than to the rest of the network. While several clustering methods exist, they split into hierarchical and non-hierarchical methods. Hierarchical methods build a hierarchy of communities by recursively dividing the network into smaller and smaller subgroups, while non-hierarchical methods directly assign nodes to communities.

For detecting the community structure, we chose the hierarchical Louvain \cite{Blondel2008} method for its speed and scalability. This algorithm runs almost in linear time on sparse graphs, therefore, it can be useful on generated test networks with increased size. It is based on modularity optimization. The modularity quotient of a network is defined by \cite{Newman2006-bw}:
\begin{equation}
Q=\frac{1}{N\av{k}}\sum\limits_{ij}\left(A_{ij}-\Gamma
\frac{k_i k_j}{N\av{k}}\right)\delta(g_i,g_j) \ .,
\end{equation}
The maximum of this value characterizes how modular a network is. Here  $A_{ij}$ is the weighted adjacency matrix, containing the admittances calculated in~\cite{HARTMANN2024101491}. Furthermore, $k_i$, $k_j$ are the weighted node degrees of $i$ and $j$ and $\delta(g_i,g_j)$ is $1$, when nodes $i$ and $j$ were found to be in the same community, or $0$ otherwise. $\Gamma$ is the resolution parameter, which allows a more generalized community detection, merging together smaller communities. 

The Hungarian power system is operated by six, territorially monopolistic companies, therefore, we chose to divide the network into six communities. To achieve this, we ran the detection algorithm starting from multiple $\Gamma$ values, we employed the divide et impera approach, starting from 0.5, we reduced the $\Gamma$ values until after 200 realizations we got lower number of communities than six. Then we divided the $\Gamma$ interval further until all realizations out of 200 realizations gave the same communities.

To reinforce the grid, similarly to our previous work~\cite{PhysRevResearch.6.013194}, we applied a simple duplication of bridges. In network analysis, a bridge (Br) refers to a link or an edge that connects nodes from different communities or components of a network. Bridges are crucial, because they establish connections between otherwise separate parts of a network, facilitating the flow of information or influence between different communities. Removing bridges can lead to a fragmentation of the network into isolated components.

Typically, community detection algorithms aim to partition nodes into distinct clusters, which naturally results in the formation of Br edges between these communities. Alternatively, one could focus on grouping edges instead. In this case, instead of connected edges, the result would be connecting or overlapping nodes. These nodes play a pivotal role in linking the various clusters of the graph. Building on this idea, we expect that in all scenarios and across the entire network, certain nodes will emerge as particularly central.

While there are several methods for link community finding, these are not always reliable and often can yield non-physical clusters. To use a more robust method, and make it comparable with our previous results, we map the graph edges to nodes and their connectivity to edges. In the literature, this is called line-graphs~\cite{harary1972linegraphs}. This way, in the transformed graph we can once again use the Louvain method, and after the inverse transform, we can find the nodes that belong to multiple communities, i.e. overlapping nodes.

\subsection{Strengthening the network via finding weak nodes - bypasses\label{sec:2D}}
Finding weak or vulnerable nodes in a network is the first step toward improving the dynamical stability of such a system. There are various approaches to identify such elements. Structurally weak elements are extracted and the network is improved as presented in Section \ref{sec:2C}. Another way to extract vulnerable elements based on dynamics is to use the local order parameter:
\begin{equation}
    R_{\text{loc.}}(t) =\langle 1 / N \sum_{j=\text{n.n}} \exp\left[i \theta_j(t)\right] \rangle,
\end{equation}
where n.n stands for nearest neighbors.

We detail the bypass-adding algorithm in one of our earlier works~\cite{PhysRevResearch.6.013194}. Shortly, the algorithm has the following logic: we compute the local Kuramoto order parameter for each node, taking into account only the first-order neighbors, and then we check the synchronization level:  $-\log_{10}{(1-\text{R}_\text{loc.})}$.

Using the above-obtained values, we split the nodes into $6$ synchronization levels. The log space is used for the splitting since generally it assures a better resolution. Notice, that the higher the '$\text{sync.}$' value, the better the synchronization.

In the next step, we determine the necessary reinforcements. For the worst-performing level, we recheck the neighborhood. If there is another weakly synchronized node as a nearest neighbor, then we will double the edge between them and the new edge will inherit the parameters of the old one. If there are only nodes belonging to a different synchronization level, then we will select the two closest ones with respect to geodesic distance and connect them. The parameters for the new link, specifically its weight, will be the average weight of the links connecting the first-order neighbors. This logic creates only new links and preserves the number of nodes.

\subsection{Strengthening the network via examining cascade sizes \label{sec:2E}}

We also simulated cascade failures, similarly as in~\cite{USAEUPowcikk} following the 'thermalization' of the system. Initial single line-cut perturbations were applied in each run, generating failure cascades as we used a fixed power threshold value: $T=0.7$. We performed 'line-cut' runs up to $t=1000 s$ with the initial cut done on each edge. Furthermore, to improve blackout simulation statistics we repeated these runs for many independent initial noise realizations.

To identify the most vulnerable links (for the ordered case and 'thermalization' with parameters $\lambda = 0.5, \alpha = 0.1, $ and $T = 0.7$), we obtained the maximum number of cascade sizes from all the realizations and sorted the values according to a descending order.

\section{Results\label{sec:3}}
This section presents the results of the three grid reinforcement methods, and provides a comparison to the traditional method used by the Hungarian transmission system operator MAVIR \cite{mavirHxE1lxF3zatfejlesztxE9siTervek}.

\subsection{Grid reinforcement with bridges\label{sec:3A}}
The networks in all scenarios were successfully divided into six communities varying the resolution parameter ($\Gamma$) values, ensuring consistency across realizations. 
The restriction of communities resulted a $\Gamma=0.28$ for scenario \#8 and $\Gamma=0.3025$ for scenario \#12.  To strengthen the grid, bridges were duplicated. This division resulted in 48 bridge links with 60 participating bridge nodes for scenario \#8 and 23 bridges involving 32 nodes for scenario \#12. For comparability of all reinforcement methods, we chose to update the networks with these numbers of links in every case for all scenarios.

On Figure \ref{fig:nets_sc8_12}(a,c) we can see the community structure and bridge linkage in both scenarios. While scenario  \#8 returns a structure closer to the real more uniform division by the energy companies, scenario \#12 gives an unequal separation with a very large community. Figure  \ref{fig:nets_sc8_12}(b) shows the bridge nodes and bypass nodes with overlapping nodes resulted from the line-graph community separation. Interestingly, not all overlapping nodes become bridge nodes.

\subsection{Grid reinforcement with bypasses\label{sec:3B}}
After performing the algorithm described in Section \ref{sec:2D} such that the resulting number of links matches the number of links obtained via bridges $\pm$ a tolerance value, typically being equal to $1$. Matching the number of new links will assure the comparability of our methods. In this spirit, we add $48$ and $23$ new links to scenarios $\#8$ and $\#12$ respectively. The initial conditions, such as the iteration number in the thermalization phase or the phase ordering can influence the local order parameter ($R_{loc}$). We generate these new links for unordered and ordered initial phases alike for the sake of completeness, but in the following, we will show the results only for the networks generated with $R_{loc}$, obtained for phase-ordered initialization. In Figure \ref{fig:nets_sc8_12}(d) we show the augmented networks (scenarios \#8 and \#12) with the bypass algorithm. We also show the nodes below the $R_{loc}$ threshold (filled red dots).

Most of the links are obtained via link doubling and not creating new ones with triangle building. Note that, the weakly synchronized nodes are focused at the center of the country in the case of scenario \#8 and they are much more scattered in the case of scenario \#12. This fact also displays the parametric (weight) differences between the networks.

\subsection{Grid reinforcement with cascades\label{sec:3C}}
\begin{figure*}[t] % [t] means place at the top of the page
    \centering
    \includegraphics[width=\textwidth]{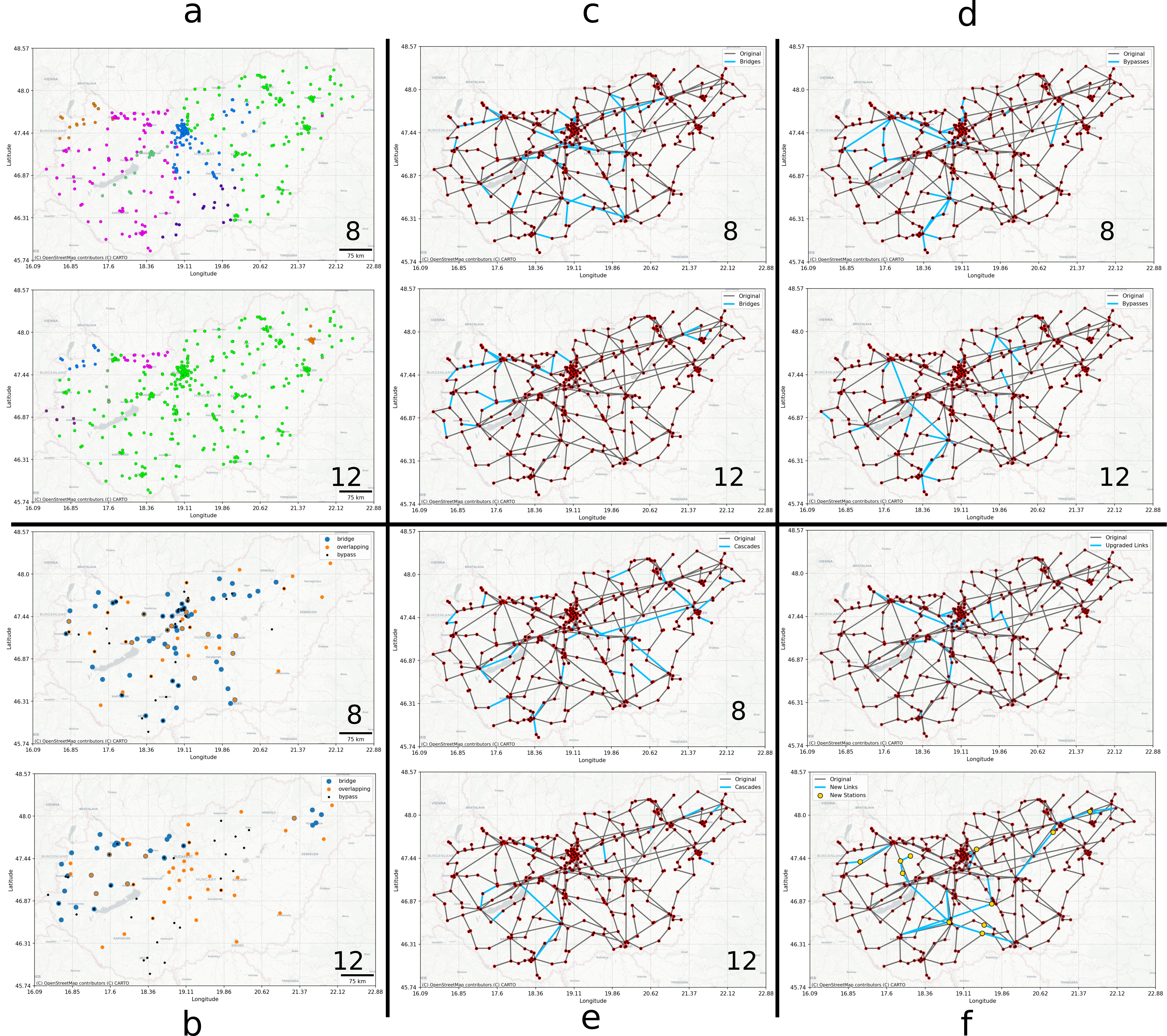} % Replace with your image file
        \caption{\justifying Top row of figures show the community structure (a) and bridge nodes vs. overlapping community nodes (b), bridge (c), bypassed (d), and cascade (e) links for scenario \#8 (upper row) and scenario \#12 (lower row). 
        a) The restricted communities with $\Gamma=0.28$ for scenario \#8 and with $\Gamma=0.3025$ for scenario \#12. b) Shows the bridge (orange) nodes and the overlapping nodes (blue points) for Scenario \#8 and \#12. The blue points are slightly larger, revealing the points that are both bridge and overlapping nodes. c) Bridge link structure for scenarios \#8 and \#12. d) scenario \#8: 48 bypass links (avg. \SI{20.5}{\km}, $37.55$ weight) and $3$ triangle-built links.
        Scenario \#12: 23 bypass links (avg. \SI{28.08}{\km}, $191.4$ weight) and $4$ triangle-built links. e) Links with maximum cascade sizes under scenario \#8 ($\ell_{ave} = 26.82$ km, $w_{ave} = 27.27$ \si{\mega\watt}) and \#12 ($\ell_{ave} = 22.072$ km, $w_{ave} = 296.57$ \si{\mega\watt}, f) grid reinforcements (upper row, 52 links) and new constructions (lower row, 37 links) planned by the Hungarian transmission system operator MAVIR.}        
    \label{fig:nets_sc8_12}
\end{figure*}

Here, we take $N = [48, 23]$ vulnerable links (for scenarios \#8 and \#12 respectively) that match the number of bridges and bypasses.The upper and lower panels of Figure~\ref{fig:nets_sc8_12}(e) show the most vulnerable links in the Hungarian power grid (in blue) under the conditions of the ordered case of scenarios \#8 and \#12.

For scenario \#8 (characterized by unique coupling strength $W_{ij}$ based on actual thermal 
capacity limits), we can observe short, mid-, and long-ranged links uniformly scattered in the territory.

On the other hand, for scenario \#12 where the link is accounted for when setting the link weights (characterized by unique coupling strength $W_{ij}$ based on actual cable admittance $Y_{ij}$ and voltage level), fewer vulnerable links can be observed. Most of which are very short-ranged such that they are unapparent on the map shown in the bottom panel of Figure~\ref{fig:nets_sc8_12}(e). The fewer number of links that experience cascades in this scenario may be due to the fact that the power linelengths are accounted for. Note that, longer cable lengths cause a larger voltage drop and more power loss.

For both scenarios we identified two vulnerable edges present in both cases: (1) 'L\'etav\'ertes - Debrecen D\'el'; and (2) 'Solym\'ar - Solt' power lines. This may suggest that these two edges are the most at risk for cascades in any configuration. However, we note that these power lines are double systems such that although they connect the same nodes, the maximum cascades have happened along any one of the cables.

For scenario \#8, the shortest link is 0.27 km, while the longest vulnerable line is 219 km long. On the other hand, for scenario \#12, fewer links can be observed with the shortest link being 0.25 km-long and the longest power line is 109 km-long.

\subsection{Comparing reinforcement methods\label{sec:3D}}

\begin{figure*}[ht] % [t] means place at the top of the page
    \centering
    \includegraphics[width=\textwidth]{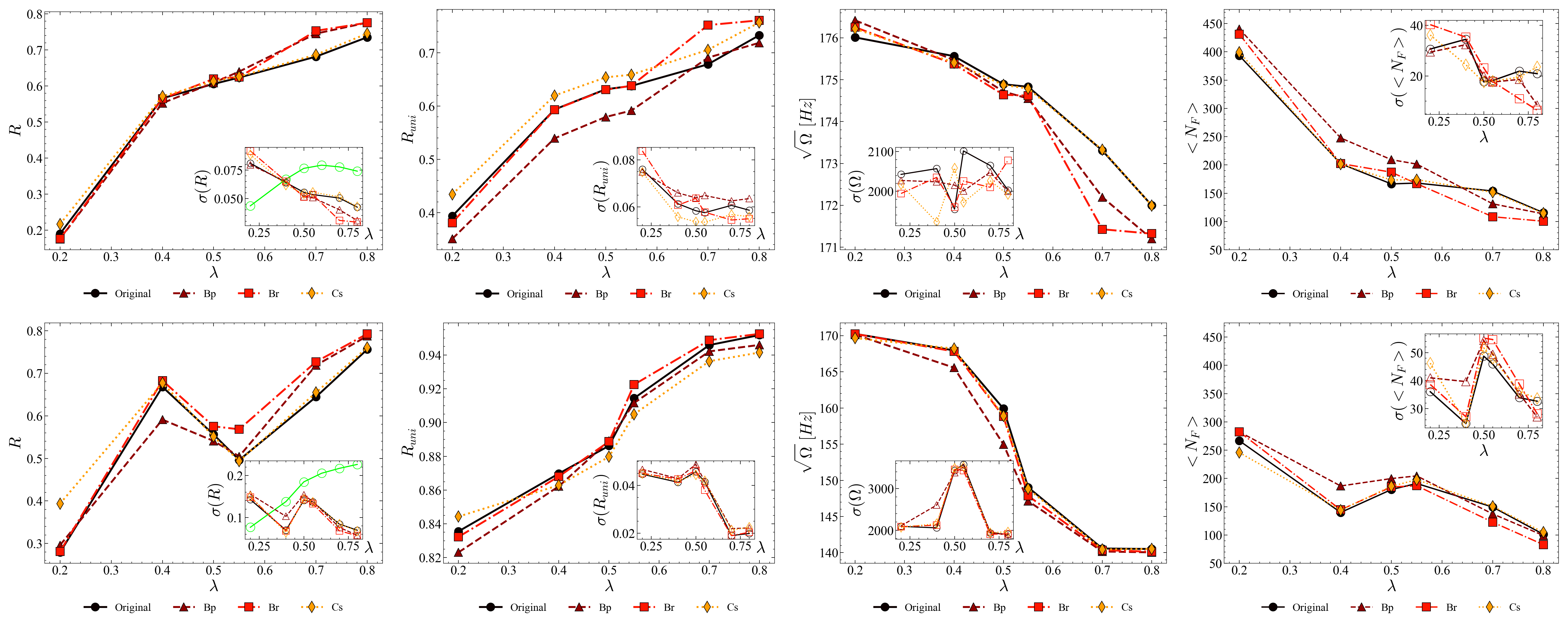} % Replace with your image file
    \caption{\justifying The Kuramoto (R), the universal order parameter ($R_{uni}$)~\cite{schroder2017}, the frequency spread ($\Omega$) in the steady state following the thermalization, and the average cascade sizes ($\langle N_F\rangle$) in the line-cut phase in function of $\lambda$ for scenario \#8 (upper row) and scenario \#12 (lower row).  The insets show the standard deviations of the corresponding quantities from the main figures in the function of $\lambda$. With a light green line, we are showing the $\sigma(R)$ values for the original network for simulations with disordered initial conditions. Using such starting states makes the double peak behavior in the standard deviation vanish in the case of scenario \#12, suggesting the possibility of having some phase-locked or frozen regions in the system.}
    \label{fig:metrics_sc8_12}
\end{figure*}

We compare the dynamic simulation results for scenarios \#8 and \#12 in Figure ~\ref{fig:metrics_sc8_12}. The best phase synchronization stabilization is obtained via bridge link reinforcements (Br) for high $\lambda$-s, while for low couplings the cascade size (Cs) method proves to be superior. The lowest frequency spreads of scenario \#12 are obtained via the (Bp) method.

Scenario \#8 shows a larger frequency spread (see the y-axis scale) for each $\lambda$, while in the case of scenario \#12 the spread gets smaller as $\lambda$ increases. The weight values of the lines in the case of scenario \#12 have typically larger values, which partially explains why the system responds better to larger global coupling.

The insets on the same figure \ref{fig:metrics_sc8_12} show the standard deviations of the corresponding quantities from the main figures in the function of $\lambda$. Where the standard deviations plotted against the control parameter exhibit a peak, a synchronization transition of the system is expected. These peaks can be found around $\lambda=0.5-0.55$ in the case of scenario \#12. It is interesting to see another increase in $\sigma(R)$ in the lower $\lambda$ region. Scenario \#8 does show such a peak at around $\lambda=0.2$.

\subsection{Characteristic results of line-cut simulations\label{sec:3E}}

\begin{figure}[t]
    \centering
    \includegraphics[width=80mm]{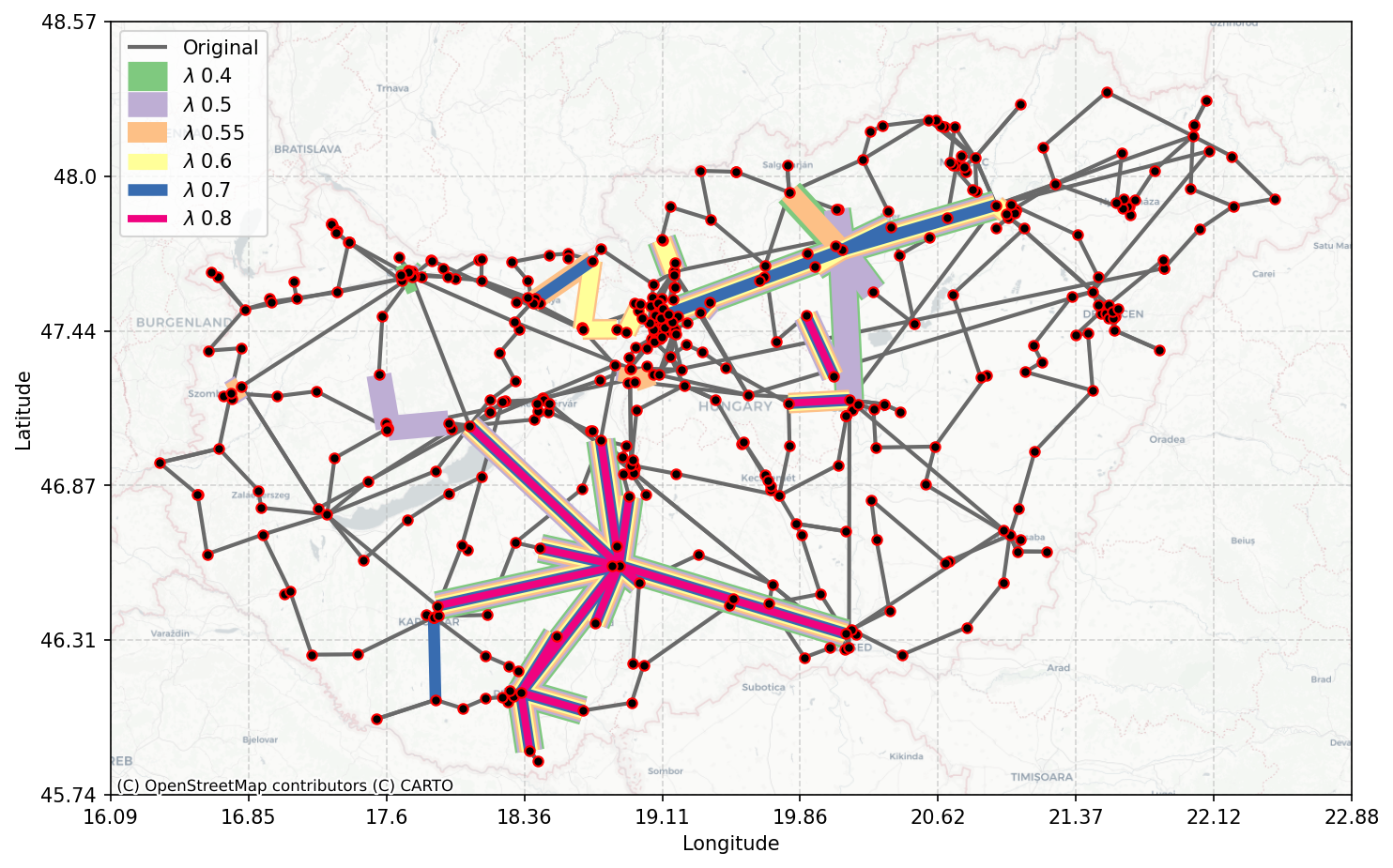}
    \caption{\justifying Map representing the links cut first. As we can see, several edges are matching for all $\lambda$. Most of the differences come from the fact that different numbers of edges will be cut in the case of different $\lambda$.}
    \label{fig:firstfails}
\end{figure}

Here we show cascade simulation results for scenario \#12 in more detail. Following $t_{max}=600 s$ thermalization time, started from phase-ordered initial conditions, there are always $N_f(t=0) \ge 3$ spontaneous line failure events for $\lambda \le 1$ at the beginning of the cascade simulations when we allowed line failures with $T=0.7$. This means that this power grid with these parameters has a small, inherent instability like in~\cite{Taher_2019}, which in real networks is stabilized using active elements~\cite{He2013,Devarapalli2022,Noco2023}. Alternatively, this could also be avoided by using $\lambda > 1.6$ couplings or high $T$-s. These most vulnerable links are typically located at dead ends near power stations, where smaller capacity dead-end nodes do not seem to swallow the input power, by following the phase differences, see Fig.~\ref{fig:firstfails}. Note, we omitted medium and low-voltage parts of the power grid and in reality, these nodes might 
distribute further the input power without active stabilization. 

\begin{figure}[t]
    \centering
    \includegraphics[width=80mm]{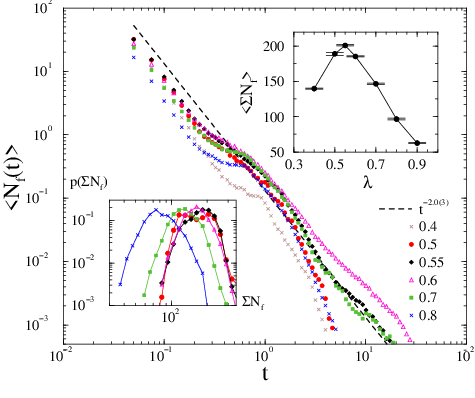}
    \caption{\justifying Time dependence of the expectation value of the line failures at a given time-step in case of scenario \#12, for different $\lambda$-s, shown by the legends. Exponential tails emerge for $\lambda \le 0.5$ and for $\lambda > 0.7$ Dashed line: PL fit for the $\lambda=0.55$ case. 
    Right inset: mean value of the total cascade sizes as the function of $\lambda$, left inset: PDF of the same. One can see a double peak
    for $\lambda=0.5$ (red bullets).}
    \label{fig:linefailures_8_12}
\end{figure}

We measured the total number of failed links, which varied between $\Sigma N_f=25$ (for $\lambda=1$) and $\Sigma N_f=200$ (for $\lambda=0.55$). This cascade size distribution dependence on $\lambda$ is shown in the right inset of Figure ~\ref{fig:linefailures_8_12}, which is obtained by sample averaging over $640 \le k \le 3000$ realizations, with different $\eta_{ik}$ quenched 'noise'.  We can see a peak at $\lambda_c(N_f) = 0.55$, which is close to the steady state fluctuation maximum of the order parameter $\lambda_c(R) = 0.5$ (see Figure ~\ref{fig:metrics_sc8_12}). Near to this coupling value: $0.55 \le \lambda \le 0.7$ we can find fat-tailed decays of the expectation values of the actual number of line failure events ($N_f(t)$) on Figure ~\ref{fig:linefailures_8_12}. This behavior breaks down for small and large $\lambda$-s and exponential decay behavior emerges. Application of least-squares PL fitting results is tail dependence, characterized by an exponent: $-2.0(3)$ at $\lambda_c$, although some weak, log-periodic oscillation, which is quite common in time-dependent solutions of the Kuramoto-model is also visible~\cite{e26121074,juhász2024finitesizescalingdynamicstwodimensional}. For $\lambda = 0.7$ a PL tail with an even smaller exponent seems to arise, reminiscent to continuously changing exponents in Griffiths phases~\cite{Griffiths}, which occur in disordered systems and was shown near to hybrid type of phase transitions~\cite{OdorSim21}. However, the effective graph dimension of this power-grid is less than 3 and the small system size allows us to speak about 'PL-like' scaling and Griffiths effects at most.

Note, that historical data~\cite{carreras2000}, DC SOC~\cite{carreras2004} and AC simulations on large continental power grids~\cite{odor2020,USAEUPowcikk} resulted in PL-tailed cascade size distributions at the critical point, but here probably small system sizes hinder to see possible PL-s. This behavior is also related to the fact that high voltage power grids tend to be small world like~\cite{USAEUPowcikk,HARTMANN2024101491} and the second order Kuramoto-model exhibits hybrid type of synchronization point~\cite{odor2023}. We can also see a series of Gaussian like peaks of PDF-s of the cascade sizes on the left inset of Figure ~\ref{fig:linefailures_8_12}, which move from the smaller to larger sizes as $\lambda$ decreases. At $\lambda=0.5$ one can observe a double peak top, corresponding to a crossover from the smaller valued peaks to the larger valued ones, which suggests a discontinuous cascade size transition.  Together with the dynamical PL tails at $\lambda_c$ we claim a manifestation of a hybrid type of transition, similarly as for the synchronization of the second-order Kuramoto-model. Results of cascade analysis for other scenarios are out of the limitations of this paper.

\section{Discussion\label{sec:4}}
Expanding the power grid demands significant investments and is intended to enhance the system's operational robustness. However, paradoxically, increasing the capacity of existing lines or installing new ones can sometimes decrease overall performance of the system due to Braess' paradox\cite{Braess1968,Braess2005}. Braess' paradox has been theoretically modeled \cite{Cohen1991,Witthaut2012,Witthaut2013,Fazlyaba2015,Nagurney2016,Coletta2016,motter2018antagonistic,Tchuisseu2018}, 
but has yet to be demonstrated in large-scale power grids.

Recent advances in topological analysis have shown some fundamental mechanisms behind Braess' paradox in power grids, offering predictive tools to identify situations where grid expansion may be disadvantageous \cite{Lacerda2021,Sch_fer_2022,Zou2024,Zhang2021,2502.09024}. In order to design resilient power infrastructures, these dynamics should be well understood, especially as intermittent renewable energy sources introduce additional variability into the grid.

To test the paradox on the different grid reinforcement results, we implemented all networks in DigSilent PowerFactory\cite{digsilentPowerFactoryDIgSILENT}. Such power system analysis software uses load-flow calculations to determine the exact loading of power lines. 
A difference compared to the Kuramoto-model is that these software also consider reactive power flows, which affects 
e.g. the loading of the lines.

\begin{figure}[H]
    \centering    \includegraphics[width=\columnwidth]{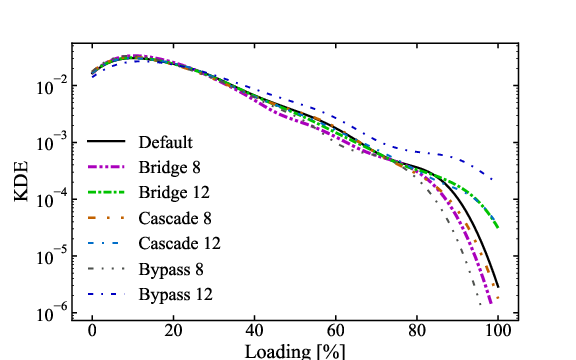}
    \caption{Kernel density estimates of the derived thermal loading for each reinforcement method, comparing it to the thermal capacity limit (black dashed 'Default' line ).}
    \label{fig:KDE_8_12}
\end{figure}

 Figure \ref{fig:KDE_8_12} shows a kernel density estimation (KDE, Eq. \ref{eq:kde}) based on the percentage loading (compared to the thermal capacity limit) of each power line. The KDE is a non-parametric method for estimating the probability density function (PDF) of a continuous random variable \cite{scott1992multivariate}. Unlike histograms, which rely heavily on correctly binning the data, KDE provides a smooth approximation of the distribution by placing a kernel (usually a Gaussian function) at each data point and summing their contributions. For calculating the KDE we used the Seaborn Python package \cite{Waskom2021}. The formula for the estimate is given by:
\begin{equation}
\hat{f}(x) = \frac{1}{n h \sqrt{2\pi}} \sum_{i=1}^{n} \exp\left(-\frac{(x - x_i)^2}{2h^2}\right),
\end{equation}\label{eq:kde}
where $\hat{f}(x)$ is the estimated probability density function, $n$ the number of data points, $h$ is the bandwidth controlling the smoothness of the function and $x_i$ are the data points. These distributions, togehter with the median values of power line loading (see Table \ref{table:scen_medians}) show that the Braess' paradox, calculated by the standard EE loading method, appears if we use the line extensions from the scenario \#12 model.
Using the line length independent model extensions of scenario \#8 we find improvements via the load-flow calculations, suggesting that the loading method might not take the lengths properly into account. 

\begin{table}[h]
    \centering
    \begin{tabular}{l|c|c}
        \hline
        Scenario & Mean & Standard deviation \\
        \hline
        Default & 20.20 & 14.81 \\
        Bridge \#8 & 18.77 & 13.67 \\
        Bridge \#12 & 19.85 & 14.69 \\
        Cascade \#8 & 19.70 & 14.71 \\
        Cascade \#12 & 20.02 & 14.80 \\
        Bypass 8 & 18.72 & 13.75 \\
        Bypass 12 & 23.24 & 17.02 \\
        \hline
    \end{tabular}
    \caption{Mean and standard deviation values of the thermal loading for different scenarios.}
    \label{table:scen_medians}
\end{table}

The results also highlight that addressing Braess' paradox in power systems requires a careful balance between expansion and optimization. While investing in the increasing of grid capacity is usually seen as the trivial way to ensure high reliability, evaluation of broader network effects is essential.

\section{Conclusions}
We have continued the Kuramoto equation-based Hungarian power grid analysis~\cite{HU387cikk} towards grid reinforcements to find the most efficient stabilization strategies. We compared the most realistic scenario (\#12) with the one, in which the power line lengths do not play a role, but the couling strengths are calculated from the maximum thermal capacities (\#8). We showed that power line lengths introduce PL admittance distributions for ~3 decades, 
similarly to our findings for the EU and USA power-grids~\cite{HARTMANN2024101491} with the exponent $1.61(3)$ 
agreeing well with the USA, EU cases, where it was $1.9(1)$.

We compared 3 different network development strategies: (Br) doubling bridges that connect different communities, (Bp) adding bypasses, around weakly synchronization nodes or doubling edges, (Cs) from which the largest cascade failures emerge. Using the same number of extra links we compared the $R(\lambda)$, $\Omega(\lambda)$, $R_{uni}(\lambda)$ and $N_f(\lambda)$ measures in the steady state.

We found that, Braess' paradox occurs in some cases, but in agreement with~\cite{PhysRevResearch.6.013194}, improvements can be achieved mostly in the intermediate $\lambda\simeq 0.5$ region, near the synchronization point $\lambda_c$ for (12) using Br (for the phases and cascade sizes) and Bp (for the frequencies). The Cs method is efficient at small $\lambda$ values in the desynchronized phase for all measures. The Cs method is related to the non-local cascade propagation unlike the others, which are more locality-related.

By looking at the some special weak lines, known from EU level simulations ~\cite{HVDCColet2024,USAEUPowcikk} it turns out that the country borders distort the results, by neglecting international loops, which enhance network redundancy. For example the Paks-S\'andorfalva connection turns out to be improvable by the Br method, the Bp and Cs simulations do not predict it to be critical, although even TSOs have known this fact for a long time. Thus, we can see limitations of country level studies, for a region embedded well in the EU power grid.

Our studies also proved that the most vulnerable links can be found at dangling ends near power-stations as huge input power cannot be dissipated by dead-end neighbors at this level of modeling.

The time dependent line failure analysis in case of scenario \#12 confirmed the hybrid type of synchronization transition, already pointed in~\cite{USAEUPowcikk}. That means for $\lambda_c$ fat tailed $N_f(t)$ distributions that can be well fitted by PLs (like in case of second order transitions), while statically we can see signatures of phase coexistence there, like in the case of first order transitions.

In our future work we plan to extend the studies related to hybrid synchronization transitions. We will also evaluate the effect of the size of the examined geographical area (i.e. simplification of cross-border connections) and the resulting grid reinforcements.

\begin{acknowledgments}
Bálint Hartmann acknowledges the support of the Bolyai János Research Scholarship of the Hungarian Academy of Sciences (BO/131/23). Kristóf Benedek acknowledges the support by the Doctoral Excellence Fellowship Programme (DCEP), founded by the National Research Development and Innovation Fund of the Ministry of Culture and Innovation and the Budapest University of Technology and Economics. Support from the Hungarian National Research, Development and Innovation Office NKFIH (K146736) is also acknowledged
\end{acknowledgments}

\section*{Data Availability Statement}
The data that supports the findings of
this study is available from the
corresponding author upon reasonable
request.

\typeout{} 
\bibliography{hetero.bib}% Produces the bibliography via BibTeX.

\newpage

\end{document}